\documentclass[aps,twocolumn,prd,epsf]{revtex4}
\usepackage{amssymb}
\usepackage{graphicx}
\usepackage{soul}
\usepackage{amsmath}
\usepackage{amsfonts}
\usepackage{comment} 
\usepackage{placeins}
\usepackage{float}
\usepackage{mathrsfs}
\usepackage{lipsum}
\newcommand{\be}{\begin{equation}}
\newcommand{\ee}{\end{equation}}
\newcommand{\bea}{\begin{eqnarray}}
\newcommand{\eea}{\end{eqnarray}}

\newcommand{\fnl}{f_{\rm NL}}

\newcommand{\di}{ {\rm d}}
\usepackage{color}

\begin{document}

\title{Non-perturbative $\delta N$ }
\author{Shailee V. Imrith}
\email{s.v.imrith@qmul.ac.uk}
\author{David J. Mulryne}
\email{d.mulryne@qmul.ac.uk}
\affiliation{\vspace{2mm}
School of Physics and Astronomy, Queen Mary University of London, Mile End Road, London, E1 4NS, UK}
\author{Arttu Rajantie }
\email{a.rajantie@ic.ac.uk}
\affiliation{\vspace{2mm}
Department of Physics, Imperial College London, London, SW7 2AZ, UK}

\begin{abstract}
We revisit the question of how to calculate correlations of the curvature perturbation, $\zeta$, 
using the $\delta N$ formalism when one cannot employ a truncated  Taylor expansion of $N$. 
This problem arises when one uses lattice simulations to probe the effects of isocurvature modes 
on models of reheating. 
Working in real space, we use an expansion in the cross-correlation between 
fields at different positions, and present simple expressions for observables such as
the power spectrum and the reduced bispectrum, $\fnl$. These take the same 
form as those of the usual $\delta N$ expressions, but with the 
derivatives of $N$ replaced by  \emph{non-perturbative $\delta N$ coefficients}. 
We test the validity of this expansion and, when compared to others in the literature, argue that our expressions are particularly well suited for use with  simulations.

\end{abstract}
\maketitle

\section{Introduction}
Inflation has been extremely successful in explaining the generation of the primordial perturbations seeding the structures of our universe, 
but the microphysics of inflation remains unknown. The simplest model consistent with existing observational data is to assume that inflaton fluctuations are solely responsible for the observed curvature perturbations. Although such a scenario is the simplest, it is
quite possible that more complicated scenarios involving additional fields, as exemplified by the curvaton model \cite{Linde:1996gt,Lyth:2001nq} and the modulated reheating model \cite{Dvali:2003em,Kofman:2003nx}, are actually realized. 
To test different inflationary theories against observations, 
one must calculate the precise form of the 
correlation functions of the primordial curvature perturbation, $\zeta$.
One technique used to do this is the separate universe approximation 
combined with the $\delta N$ formalism \cite{Lyth:1984gv,Wands:2000dp,Sasaki:1995aw,Lyth:2005fi,Lyth:2004gb} . 
In this approach, $\zeta$ is given by the 
perturbation in the local e-folding number
\begin{equation}
\zeta (\mathbf{x}) = \delta N( \mathbf{x}) = N(\vec{\chi} (\mathbf{x})) - \bar N  \,,
\label{dN}
\end{equation}
where $N$ is the number of e-folds between an initial flat hypersurface at some early time (such as horizon crossing) and a final uniform density 
hypersurface at some later time (such as the end of inflation or after reheating), and  $\bar N=  \langle N \rangle$. Throughout, angle brackets indicate an ensemble average.  We consider $n$ fields labeled, $\chi^I$, where 
$I$ runs from $1$ to $n$, and for convenience we introduce the vector, $\vec{\chi}$, where each element 
represents one of the $n$ fields.

$N$ is 
calculated by assuming that locally the universe can be approximated as a Friedmann-Robertson-Walker spacetime, 
and hence is a function of the local 
field values on the initial flat hypersurface. Standard practice is to approximate $\delta N$ by making a Taylor 
 expansion in the initial field values and keeping only 
 a small number of terms. In some cases, however, 
 $N$ depends very sensitively on the initial field values, and 
 a truncated Taylor expansion is not a good approximation.  Such cases include 
those in which a light field in addition to the inflaton influences the dynamics of 
non-perturbative reheating \cite{Chambers:2007se,Chambers:2008gu,Chambers:2009ki,Bond:2009xx}. In this paper we return to the issue of how to deal with such cases. As we will see, an alternative expansion is sometimes possible. 
 
Although the primary motivation for our work is the interpretation of the results of lattice simulations, here we study the question generally. 
Our approach employs many of the key ideas contained in the work of Suyama and Yokoyama \cite{Suyama:2013dqa}, 
and our results are broadly 
equivalent to theirs. In that work, however, a key step was to make a Fourier transform of the $N$ function (treated as 
a function of a single field value). This is useful 
for analytic manipulations, but leads to expressions for the correlation functions that are less useful if an exact form for 
$N$ is unknown, or, as can be the case for lattice simulations, it is not efficient even to  
calculate the form of the $N$ function explicitly. The expressions we arrive at are more applicable in this setting, lending themselves 
to a Monte Carlo approach,  a 
point we return to later. 
Our methods are more closely related to the work 
of Bethke, Figueroa and Rajantie \cite{Bethke:2013aba,Bethke:2013vca} who considered the 
power spectrum of gravitational waves from massless 
preheating, though depart from both these earlier studies by considering $n$ fields whose initial 
probability distribution need not be precisely Gaussian. We perform explicit calculations 
only for the two and three-point functions of $\zeta$, but the method extends trivially to higher point functions. For other related work with a different approach to ours, see \cite{Vennin:2015hra} and \cite{Vennin:2016wnk} where the authors develop and apply a non-perturbative formulation of $\delta N$ by incorporating the stochastic corrections to $N$. 

The remainder of this paper is structured as follows: In section \ref{nonpert}, we develop and describe the non-perturbative $\delta N$ formalism. Our main results are presented in section \ref{expandedmethod}. We then apply this formalism in section \ref{examples} to both analytic and non-analytic examples and make useful comparisons to regular $\delta N$ formalism. Finally, we conclude in section \ref{conclusion}.

\section{Non-perturbative $\delta$N Formalism}
\label{nonpert}

\subsection{Regular $\delta N$}
\label{regular}
In the standard $\delta N$ approach, to calculate the 
correlations of $\zeta$ in Fourier space one first assumes that 
the statistical distribution of field space perturbations 
is known on the initial flat hypersurface. The field perturbations, $\delta \chi^I = \chi^I - \bar \chi^I$,  are taken to be close to Gaussian with the power spectrum defined as 
\begin{equation}
\label{fieldSpect}
\langle \delta \chi^I_\mathbf{k_1} \delta \chi^J_\mathbf{k_2} \rangle = (2 \pi)^3 \Sigma^{IJ} (k) \delta^3(\mathbf{k_1} + \mathbf{k_2})\,.
\end{equation}
Higher 
order cumulants are either taken to be completely negligible, 
or are included in 
the formalism, order by order, considering first the three-point function 
on the initial hypersurface,
\be
\label{field3pt}
\langle \delta \chi^I_\mathbf{k_1}  \hspace{-.6mm} \delta \chi^J_\mathbf{k_2}\hspace{-.6mm} \delta \chi^K_\mathbf{k_3} \rangle \hspace{-.5mm}= \hspace{-.5mm} (2 \pi)^3 \alpha^{I\hspace{-.5mm} J\hspace{-.5mm}K} (k_1,k_2,k_3) \delta^3(\mathbf{k_1}\hspace{-.5mm} + \hspace{-.4mm}\mathbf{k_2}\hspace{-.5mm}+\hspace{-.4mm}\mathbf{k_3}) \,,
\ee
and then successive higher order cumulants. 
To utilise Eq.~(\ref{dN}), one first makes a 
Taylor expansion of the $N$ function in terms of 
$\delta \chi^I(\mathbf{x})$, such that to second order 
\be
\delta N(\mathbf{x})  \hspace{-.5mm} = \hspace{-.5mm} N_{\hspace{-.5mm} ,I} \delta \chi^I \hspace{-.5mm} (\mathbf{x})  +\hspace{-.3mm}  \frac{1}{2}N_{\hspace{-.5mm} ,I\hspace{-.5mm} J} \hspace{-.3mm} \left( \delta \chi^I \hspace{-.5mm} (\mathbf{x})\delta \chi^J \hspace{-.5mm} (\mathbf{x}) \hspace{-.5mm} - \hspace{-.5mm} \overline{\delta \chi^I \delta \chi^J} \right ),
\label{dN2}
\ee
where  $\overline{\delta \chi^I \delta \chi^J} = \langle \delta \chi^I(\mathbf{x}) \delta \chi^J(\mathbf{x})\rangle$.
One then considers the Fourier transform of Eq.~(\ref{dN2}),
and  forms the desired correlation of $\zeta(k)$, typically  keeping only the leading terms.
Finally, applying a Wick expansion, and 
using Eq.~(\ref{fieldSpect}) and any non-zero higher order cumulants, one produces an 
expression for the Fourier space 
correlations of $\zeta$ at the final time in terms of the 
correlations of the fields at the early time.  
For example, the two and three-point functions of $\zeta$, defined in 
terms of the power spectrum $P_\zeta$ and bispectrum $B_\zeta$ are given by
\begin{eqnarray}
\langle \zeta_\mathbf{k_1} \zeta_\mathbf{k_2} \rangle &\equiv& (2\pi)^3 P_\zeta(k_1) \delta^3(\mathbf{k_1} + \mathbf{k_2}) \nonumber \\[.7ex]
 &=& (2\pi)^3 N_{,I} N_{,J} \Sigma^{IJ}(k_1) \delta^3 (\mathbf{k_1} + \mathbf{k_2}) \label{twoStandard} \\[1ex]
\langle \zeta_\mathbf{k_1} \zeta_\mathbf{k_2}   \zeta_\mathbf{k_3} \rangle &\equiv& (2\pi)^3 B_\zeta(k_1,k_2,k_3) \delta^3(\mathbf{k_1} + \mathbf{k_2}+ \mathbf{k_3}) \nonumber \\[.7ex]
 &=& (2\pi)^3 \Big [ N_{,I} N_{,J} N_{,K} \alpha^{I\hspace{-0.3mm}J\hspace{-0.3mm}K}(k_1,k_2,k_3) \nonumber \\[.2ex]
 &~&+ \, \big(N_{,I}N_{,J}N_{,KL} \Sigma^{I\hspace{-0.3mm}K}(k_1)\Sigma^{J\hspace{-0.3mm}L}(k_2)
 \nonumber \\[.2ex] 
 &~& + \,{\rm cyclic} \big )  \Big ] \delta^3(\mathbf{k_1} + \mathbf{k_2} + \mathbf{k_3}) \,.
 \label{threeStandard}
\end{eqnarray}
We note that here and throughout when we discuss correlations of fields we always mean those 
at the initial time, and when we discuss correlations of $\zeta$ 
we always mean those at the final time. 
Finally we also note that taking $\alpha(k_1,k_2,k_3)$ to be zero (along with higher order cumulants) is a good approximation 
for canonical theories with the field statistics evaluated at horizon crossing, but not otherwise.

\subsection{$\delta N$ without a Taylor expansion}

\subsubsection{Preliminaries and notation}

We will now consider how to proceed if $N$ is not well approximated by a Taylor expansion. 
In this case, it proves convenient to stay in real space and calculate the correlations of $\zeta$ there, including 
information from all scales, and 
only then to Fourier transform the correlation (for each of the spatial coordinates which appear) 
to calculate the Fourier space
correlations over observational scales or equivalently to coarse-grain the correlations over these scales. 
This procedure is most convenient because 
$N$ is a function of the fields which are in turn a function of spatial position. One could attempt to 
treat $N(\vec{\chi}(\mathbf{x}))$ as a function of $\mathbf{x}$ and 
Fourier transform it directly, but given that it is a non-linear function of the fields, the result 
would not be a simple function of the Fourier coefficients of the fields, $\vec{\chi}(k)$, which are the objects 
we have information about. 
 
For later convenience, therefore, let us introduce some notation for 
the statistics of the field space perturbations in real space as
\begin{equation}
\label{realSigR}
\langle \delta \chi^I(\mathbf{x}_1) \delta \chi^J(\mathbf{x}_2) \rangle =  \Sigma^{IJ} ({r_{12}})\,,
\end{equation}
where ${r_{12}} = |\mathbf{x}_1 - \mathbf{x}_2|$, and
\begin{equation}
\label{realAlphaR}
\langle \delta \chi^I(\mathbf{x}_1) \delta \chi^J(\mathbf{x_2}) \delta \chi^K(\mathbf{x_3}) \rangle = \alpha^{IJK}({r_{12}},{r_{23}},{r_{31}})\,.
\end{equation}
In an abuse of notation we use the same symbol for the correlations as for the related objects in Fourier space (defined in 
Eq.~(\ref{fieldSpect}) and Eq.~(\ref{field3pt})), but it 
will always be clear from the context which we mean. We further define the shorthand notation
\begin{eqnarray}
 \langle \delta \chi^I(\mathbf{x}_1) \delta \chi^J(\mathbf{x}_1) \rangle \hspace{-1mm}&= &\hspace{-1mm} \Sigma^{IJ} \label{realSig} \\[.5ex]
\langle \delta \chi^I(\mathbf{x}_1) \delta \chi^J(\mathbf{x}_1)\delta \chi^K(\mathbf{x}_1) \rangle \hspace{-1mm} &=& \hspace{-1mm}\alpha^{IJK}\,,
\label{realAlpha}
\end{eqnarray}
since when evaluated at the same spatial position the 
correlations are no longer functions of space.

Finally, we introduce more short hand 
 notation such that the evaluation of a function at a
given spatial
position is denoted using a subscript, for example
$\zeta_1 =\zeta(\mathbf{x}_1)$, $\chi^I_1 = \chi^I(\mathbf{x}_1)$
 and  $N_1 = N(\vec{\chi}(\mathbf{x}_1))$. This is 
 helpful to keep our expressions to a manageable size when we are 
 considering many spatial positions in one expression.

\subsubsection{A non-perturbative expression}
\label{anpexp}

When $\zeta$ cannot be written in terms of an expansion in $\delta \chi^I(\mathbf{x})$, one cannot write the 
correlations of $\zeta$ in 
terms of a finite number of correlations of the field perturbations. Instead one must 
fall back on the definition of the ensemble average, and write the $m$-point function, 
$\langle \zeta(\mathbf{x}_1) ... \zeta(\mathbf{x}_m) \rangle$, 
in terms of the full $n\times m$ joint probability distribution for the $n$ fields 
evaluated at the $m$ spatial positions. This is given as

\begin{eqnarray}
\langle \zeta_1 ... \zeta_m \rangle  &=& \langle (N_1-\bar N) ... (N_m - \bar N) \rangle  \nonumber \\ 
&=& \int {\rm d} \vec{\chi}_1 \, ...  \int {\rm d} \vec{\chi}_m (N_1-\bar N) ... (N_m - \bar N) \nonumber \\
&~& ~~~~~~~~~ ~~~~~~~~~~\times {\cal P}(\vec{\chi}_1, ..., \vec{\chi}_m)\,,\label{corrFull}
\end{eqnarray}
where ${\cal P}$ is the joint probability distribution for the $m \times n$ variables $\chi^I_i$, 
and we have used the subscript notation defined at the end of the previous subsection.
 The integral is over all the fields evaluated at the $m$ distinct spatial positions. 
If $N$ is a simple function,  and if ${\cal P}$ can be taken to be Gaussian, which is often a very good approximation, 
then it is possible to evaluate Eq.~(\ref{corrFull}) analytically. More generally 
it is possible to evaluate it numerically. We will see examples of both for the single field case 
in section (\ref{examples}).

Although not presented explicitly there,  Eq.~(\ref{corrFull}) in the single field case is the
starting point for the work of Suyama and Yokoyama \cite{Suyama:2013dqa}. In that work the focus 
is on extracting analytic results for the moments of $\zeta$ when an analytic form for $N$ is known. They proceed by assuming that 
the probability distribution is exactly Gaussian, and by considering the Fourier transform of the $N(\chi)$ function (when $N$ is 
treated as function of $\chi$). 
In this case general expressions for the correlations of $N$ are known in terms of the Fourier coefficients of 
$N$ and the variance of $\chi$ (these are given in Eq.~(9) of Ref~\cite{Suyama:2013dqa}), 
and they proceed to work directly with these 
expressions in their paper. In 
our work we work directly with Eq.~(\ref{corrFull}). This more direct route still 
allows  Eq.~(\ref{corrFull}) to be evaluated analytically 
for specific forms of the $N(\chi)$ function, 
but also allows us to introduce additional fields, to expand the distribution, and to consider non-Gaussian initial conditions 
in a  straightforward manner.

\subsection{Expansions of the probability distribution}

While it is possible to work directly with Eq.~(\ref{corrFull}), it is rather cumbersome in practice, especially 
if it needs to be integrated numerically or if the probability distribution, ${\cal P}$, cannot be taken to be Gaussian. 
Moreover,  if a numerical evaluation is needed the process becomes particularly involved when 
the correlations are converted to Fourier space, to calculate observable quantities such as the power spectrum and bispectrum on
observable scales. 
In this case one must Fourier transform the real space 
correlations in each of the $m$ spatial coordinates that appear, which requires that the integral, Eq.~(\ref{corrFull}), is 
evaluated first at a sufficient number of points 
in real space and then transformed to Fourier space.

\subsubsection{Two expansions}

Thankfully, for many applications there is still an approximate method available 
even when $N$ cannot be Taylor expanded.
Rather than expanding the 
$N$ function, the idea is to employ, instead, expansions of the distribution ${\cal P}$.

First ${\cal P}$ is expanded around a Gaussian distribution 
employing a Gauss-Hermite expansion. 
In the inflationary context 
a Gauss-Hermite expansion for the distribution of field perturbations 
was used by Mulryne {\it et al.} \cite{Mulryne:2010rp}, and 
is justified since the field perturbations produced by inflation are very close 
to Gaussian \cite{Hawking:1982cz,Hawking:1982my,Bardeen:1983qw,Guth:1985ya,Maldacena:2002vr,Seery:2005wm,Seery:2005gb,Chen:2006nt} (even for levels of non-Gaussianity far in excess of observational bounds). 

Next this distribution is expanded in the 
cross correlation between fields evaluated at different spatial positions, $\Sigma^{IJ}({r}_{ij})$  
with $i\neq j$, around the distribution for the field perturbations evaluated at the same 
spatial position, i.e, we assume that $\Sigma^{IJ}({r}_{ij}) < \Sigma^{IJ}$ (recall $\Sigma^{IJ}\equiv \Sigma^{IJ}(\mathbf{0})$).
This expansion has been utilised previously 
by Suyama and Yokoyama \cite{Suyama:2013dqa}
and by Bethke {\it et al.} \cite{Bethke:2013aba,Bethke:2013vca}. 
It is at least partially justified if 
the power spectrum for the field fluctuations $\delta \chi^I(\mathbf{k}_1)$ is close to scale invariant, since 
then for two positions, $\mathbf{x}_1$ and $\mathbf{x}_2$, separated by 
a distance close to the size of the observable universe we 
find that $\Sigma^{IJ}(\mathbf{r}_{12}) $ is roughly two orders of magnitude smaller than $\Sigma^{I J}$. 
We will always be interested either in real space correlations of $\zeta$ coarse-grained on these large observationally relevant 
scales, or equivalently in the Fourier space correlations for small wavenumbers. 
See, however, \S~\ref{limitations} for caveats and a more detailed discussion.

\subsubsection{An interlude on our expansions}
\label{expansions}

Let us begin in the abstract, before moving to the inflationary context, and consider the 
distribution for a set of close to Gaussian 
coupled variables $y_\alpha$ denoted by the vector $\mathbf{y}$. This is given by the 
Gauss-Hermite expansion,
\be
\label{GH}
\hspace{-.003cm} {\cal P}(\mathbf{y}) \hspace{-.7mm}= \hspace{-.7mm}{\cal P}_{_{\rm G}} (\mathbf{y})\left (\hspace{-.6mm} 1 \hspace{-.7mm} + \hspace{-.7mm}  \frac{ A^{-1}_{\alpha \epsilon} A^{-1}_{\beta \eta} A^{-1}_{\gamma \mu}   \alpha_{\epsilon \eta \mu} H_{\alpha \beta \gamma}(\mathbf{z})}{6} \hspace{-.7mm}  + \hspace{-.7mm} \,...  \right ), \hspace{-.165cm} 
\ee
where the subscript ${\rm G}$ indicates a multivariate Gaussian distribution with covariance matrix $\Sigma_{\alpha \beta} \equiv \langle  \delta y_\alpha  \delta y_\beta   \rangle = A_{\alpha \epsilon} A_{\beta \epsilon}$, and where $\alpha_{\alpha \beta \gamma}  \equiv \langle  \delta y_\alpha  \delta y_\beta  \delta y_\gamma  \rangle$. Where $\delta y_\alpha = y_\alpha - \bar y_\alpha$ and $\mathbf{z}$ is the vector with elements $A^{-1}_{\alpha \beta} \delta y_\beta$. The 
functions in the expansion are products of Hermite polynomials
defined by a generalised version of Rodrigues' formula, such that   
$H_{\alpha  \beta \gamma} = - \partial^n/\partial z_\alpha \partial z_\beta \partial z_\gamma \exp(-\mathbf{z}^2)$.
We will only need the result that $H_{\alpha \beta \gamma}({\bf z}) =   \delta y_\alpha \delta y_\beta \delta y_\gamma$ if  $\alpha \neq \beta \neq \gamma$.
A multivariate Gauss-Hermite 
expansion around a Gaussian distribution has been employed elsewhere in the cosmological 
literature for various purposes (see, for example, \cite{Mulryne:2010rp,Mulryne:2009kh,Contaldi:2001wr,Juszkiewicz:1993vk,Matarrese:2000iz,Amendola:2001up,Amendola:1996ny,Seery:2006wk,Watts:2002rz}).

Now let us consider the second expansion we will need to make. We note that 
if any of the elements of the variance matrix $\Sigma_{\alpha \beta}$ 
are small in the sense that we can neglect terms involving their square,
it is possible to make a Taylor expansion of the distribution, Eq.~(\ref{GH}), in this element. For 
our purposes to make use of such an expansion, we will only need the following results
\begin{eqnarray}
\frac{\partial  {\cal P}_{_{\rm G}}}{\partial  \Sigma_{\alpha \beta}}\hspace{-0.05cm}&=&\hspace{-0.05cm} \frac{1}{2} {\cal P}_{_{\rm G} }  \delta y_\gamma \delta  y_\delta   
\Sigma^{-1}_{\gamma \alpha } \Sigma^{-1}_{\delta \beta},
 \label{expand1}\\[2ex]
\hspace{-0.3cm}\frac{\partial^2  {\cal P}_{_{\rm G}}}{\partial  \Sigma_{\alpha \beta} \partial \Sigma_{\gamma \delta}} \hspace{-0.1cm}&\supset& \hspace{-0.1cm}  \frac{1}{4}  \delta  y_\epsilon \delta y_\eta \delta y_\mu \delta y_\nu {\cal P}_{_G} \Sigma^{-1}_{\alpha \epsilon} \Sigma^{-1}_{\beta \eta }  \Sigma^{-1}_{\gamma \mu} \Sigma^{-1}_{\delta \nu}.
\label{expand2}
\end{eqnarray}
In this context $ A\supset B$ denotes that A contains B as well as some other terms. 

\subsubsection{Calculating correlations of $\zeta$ using the expansions}
\label{expandedmethod}

Finally, we can use these expansions in the context at hand. We assume that the distribution 
which appears in Eq.~(\ref{corrFull}) 
for the $m\times n$ independent variables, $\chi^I_i$, is both close to Gaussian, so that 
the Gauss-Hermite expansion can be employed, 
and moreover that the $n\times m$ variate Gaussian 
which appears in this expansion can be further expanded in the cross-correlations $\Sigma^{IJ}(r_{ij})$ where $r_{ij} \neq 0$. 
Specialising to the two-point function and employing Eq.~(\ref{corrFull}) with both expansions, one finds that at 
leading order
\begin{widetext}
\begin{eqnarray}
\int {\rm d}  \vec{\chi}_1  {\rm d} \vec{\chi}_2 \, {\cal P} (\vec{\chi}_1,\vec{\chi}_2 )  (N_1-\bar N) (N_2-\bar N) 
  \approx \,   \Sigma^{IJ}(r_{12})  \Sigma^{-1}_{IK} \Sigma^{-1}_{JM} \hspace{-0mm} \int\hspace{-0mm} {\rm d} \vec{\chi}_1 {\cal P}_{_{\rm G}}(\vec{\chi}_1) 
 \delta \chi^K_1(N_1 -\bar N) \times  \int {\rm d}\vec{\chi}_2 {\cal P}_{_{\rm G}}(\vec{\chi}_2) \delta \chi^M_2 (N_2- \bar N)\;\;\;
\label{twoExpand}
\end{eqnarray}
\end{widetext}
where  $\Sigma^{-1}_{IJ}$ is the inverse of $\Sigma^{IJ}$, which for clarity we 
recall is the covariance matrix of field perturbations evaluated 
at the same point in real space. This leading term comes from the first order term 
in the cross-correlation Taylor expansion, which is calculated from Eq.~(\ref{expand1}). There is 
no contribution from the zeroth order term because one  
needs at least 
one $\delta \chi_i$ to accompany 
each $N_i $ function so that the expectation of a given term isn't zero.
Note that the Gaussian probability distribution which appears twice on the right hand side of this expression is 
the $n$ dimensional distribution for fields evaluated at only a single position, and we have retained both the subscripts 
$1$ and $2$  only for clarity as to how the expression arises.
We can write Eq.~(\ref{twoExpand}) as 
\be
\langle \zeta_1 \zeta_2 \rangle \approx \tilde N_I \tilde N_J \Sigma^{IJ}(r_{12}),
\label{dNnew}
\ee
where we have defined
\be
\tilde N_I = \Sigma^{-1}_{IJ}  \int \hspace{-.2mm} {\rm d} \vec{\chi}_1\hspace{.6mm}  {\cal P}_{_{\rm G}}(\vec{\chi}_1) N_1 \hspace{-.1mm}
\delta \chi^J_1,
\ee
which is analogous to the first derivative of $N$ used in Eq.~(\ref{dN2}). The 
spatial position indicated by the subscript $1$ is of course arbitrary.

Following the same procedure for the three-point function 
one finds that we must keep two terms at leading order, one involves the $\alpha$ term from the 
Gauss-Hermite expansion, and the second is second order in the cross-correlation expansion 
and arises from the term given in Eq.~(\ref{expand2}). These are the first terms to contribute  
since again we need at least one $\delta \chi_i$  to accompany each of the three  $N_i $ functions in 
the three-point function 
so that the expectation value of a given term is not zero. 
One finds 
\begin{eqnarray}
\langle \zeta_1 \zeta_2 \zeta_3 \rangle &\approx& \tilde N_I \tilde N_J \tilde N_K \alpha^{IJK}(r_{12},r_{23},r_{31}) \nonumber \\
&~&+\,\big (\tilde N_I \tilde N_J \tilde N_{KL} \Sigma^{IK}(r_{12})\Sigma^{JL}(r_{23})\nonumber \\ 
&~&~\,\,\,\,\,+ {\rm cyclic} \,\big)
\label{dNnew2}
\end{eqnarray}
where we have defined
\be
\tilde N_{I\hspace{-.4mm} J} \hspace{-.6mm}=\hspace{-.6mm} \Sigma^{-1}_{I\hspace{-.2mm} K}\Sigma^{-1}_{J\hspace{-.2mm} L} \hspace{-1.2mm}  \int \hspace{-1.2mm}{\rm d} \vec{\chi}_1\hspace{.6mm}  {\cal P}_{_{\rm G}}(\vec{\chi}_1) (N_1- \bar N) \hspace{-.1mm} \delta \chi^K_1 \delta \chi^L_1 
\ee
analogous to the second derivative of $N$ used in Eq.~(\ref{dN2}).

Using these expressions, and 
accounting for only the second term of Eq.~(\ref{dNnew2}), the local contribution to the reduced bispectrum $\fnl$, takes the 
famous form
\be
\frac{6}{5}\fnl = \frac{\tilde N_I \tilde N_{IJ} \tilde N_J}{(\tilde N_K \tilde N_K  )^2}\,.
\label{fnl}
\ee

It is important to note that Eqs.~(\ref{dNnew}) and (\ref{dNnew2}) combined with the definition of $\tilde N_I$
and $\tilde N_{IJ}$ 
represent a significant simplification, since the spatial dependence of 
the two-point function of $\zeta$ is defined entirely through that of the 
field fluctuations. This is an important advantage, particularly if the correlation of $\zeta$ is 
to be evaluated numerically, since otherwise the numerics would need to be repeated for 
many values of $r_{12}$, while in this case $\tilde N_I$ and $\tilde N_{IJ}$ need only be evaluated once. This allows us 
to pass immediately to Fourier space, and to write the power spectrum and bispectrum of $\zeta$ as
\begin{eqnarray}
P_\zeta(k) &\approx& \tilde N_I \tilde N_J \Sigma^{IJ}(k)\\[1ex]
B_\zeta(k_1,k_2,k_3) &\approx& \tilde N_I \tilde N_J \tilde N_K \alpha^{I\hspace{-0.3mm}J\hspace{-0.3mm}K}(k_1,k_2,k_3) \nonumber \\[.5ex]
 &~& +\,\big( \tilde N_{I} \tilde N_{J} \tilde N_{KL}  \Sigma^{I\hspace{-0.3mm}K}(k_1)\Sigma^{J\hspace{-0.3mm}L}(k_2) \nonumber \\[.5ex]
 &~& ~~~~~~~~~~~~~~~~~~+ {\rm cyclic} \,\big ). ~
\end{eqnarray}

\subsubsection{Further simplifications for typical applications}
\label{simple}

A further simplification occurs if we assume that the field fluctuations 
are uncorrelated such that $\Sigma^{IJ}$ is diagonal. The simplest case is 
if all fields have the same variance, such that 
\begin{equation}
\label{four}
\Sigma^{IJ} = \delta^{IJ} {P_\chi} \,
\end{equation}
which is a good approximation at horizon crossing during 
inflation. 
More generally the covariance matrix might be diagonal but 
with different entries, such that 
\begin{equation}
\label{four}
\Sigma^{IJ} =  \delta^{IJ} P_{\chi^I}\, 
\end{equation}
where no summation is implied. This would be the case in a model with one inflaton field and a set of 
fields that were purely isocurvature modes during inflation. 
In this case one finds $\tilde N_I$ simplifies  to 
\begin{eqnarray}
\tilde N_I &=&  \frac{1}{P_{\chi^I}} \int {\rm d}  \vec{\chi}_1  {\cal P}_{_{\rm G}}(\vec \chi_1) N_1 \delta \chi^I_1 \\[1ex]
               &\equiv& \frac{1}{P_{\chi^I}} \langle \delta \chi_1^I N_1\rangle_{_{\rm G}} \label{NI}	
\end{eqnarray}
and
$\tilde N_{IJ}$ simplifies to
\begin{eqnarray}
\tilde N_{IJ} \hspace{-.1cm}&=& \hspace{-.1cm} \frac{1}{P_{\chi^I}P_{ \chi^J}} \hspace{-.1cm}\int\hspace{-.1cm} {\rm d}  \vec{\chi}_1   {\cal P}_{_{\rm G}}(\vec \chi_1)( N_1 - \bar N) \delta \chi^I_1 \delta \chi^J_1 \\[1ex]
               &\equiv& \frac{1}{P_{\chi^I}P_{\chi^J}} \langle \delta \chi^J_1 \delta \chi_1^I (N_1-\bar N) \rangle_{_{\rm G}} \label{NIJ}	
\end{eqnarray}
because in this case, the covariance matrix is diagonal, 
${\cal P}_{_{\rm G}}(\vec\chi_1) = \prod_I{\cal P}_{_{\rm  UG}}(\chi^I_1)$, where  subscript ${\rm  UG}$ 
now stands for a univariate Gaussian.

\subsubsection{A Monte Carlo approach}
\label{monte}

In the paper,  the examples we consider will be of cases where there is a known $N$ function, either 
an analytic one, or one that has been calculated numerically. 
When we utilise the simplified expressions given 
above, we will therefore use the known $N$ function and integrate 
Eqs.~(\ref{NI}) and (\ref{NIJ}), either analytically 
or using numerical methods.
However, a major motivation 
of our work is to allow the future study of cases in which it may not be desirable to first calculate  
$N$ as a function of the initial field values. We defer doing this to future work, but it is 
worth laying out a case for the suitability of our expressions for this purpose. 
It may be that the $N$ function is highly featured, such as 
in the case of massless preheating \cite{Chambers:2007se,Chambers:2008gu,Bond:2009xx,Kohri:2009ac,Prokopec:1996rr,Greene:1997fu}, and that 
first calculating the function accurately may not be the most 
efficient path to accurately evaluating $\tilde N_I$ and $\tilde N_{IJ}$. Instead one might choose to adopt 
a Monte Carlo approach, in which values of the initial field(s) $\chi^I$ 
are drawn from a Gaussian distribution, and for each draw $N$ is evaluated numerically. $\tilde N_I$, for example, is then calculated 
by evaluating  $\delta \chi^I N$ for each draw, and the values summed and divided by the number of draws. The convergence 
of the result can be monitored. This was the approach adopted in the gravitational wave case by Bethke  { \it et al.} \cite{Bethke:2013aba,Bethke:2013vca}. In contrast to previous work \cite{Suyama:2013dqa}, our 
expressions are ideal for this purpose.

\subsubsection{Limitations}
\label{limitations}

Subsections \ref{expandedmethod} and \ref{simple} represent the main results of our paper. 
In section \ref{examples} we will see them in practice, and test their validity. First, however, let us consider 
what we expect to be their limitations in terms of the approximations we have employed.

The first limitation stems from the fact that we expand the probability distribution in the cross correlations 
between distinct spatial positions, and then integrate to calculate the correlations of $\zeta$. 
This means that the resulting expansion  is not guaranteed to be a good one (in the sense that it will converge), 
even if the expansion of the probability distribution does converge. So while $\Sigma(r_{ij}) \ll \Sigma$ is sufficient 
for the probability expansion to be valid, this is not sufficient for the correlations calculated from it to 
converge. This effectively means that we have to test the 
validity of our expressions on a case by case basis. 

The second related issue comes from the fact that even if the series does converge, there 
is no guarantee that the leading term in the cross correlations is sufficient. An extreme example 
follows from the fact that it is possible for the ``leading" term we quote 
above to be zero.
For the two-point function this occurs when the $N$ function is symmetric in one of fields (about $\bar \chi$) -- an even 
function in the single field case. In this case, considering Eq.~(\ref{NI}) for a single field, 
we see that $\langle \delta \chi N(\chi) \rangle_{_{\rm G}}  =0$.
Although realistic functions of $N$ will never be fully even or fully odd, this issue 
should be  borne in mind.

In both cases one thing that can be done is to check that the sub-leading term is subdominant to 
the leading term. Although not proof of convergence this is a simple way to check that the method is 
working as intended. 
For example, in the single field case where the sourcing scalar field is Gaussian, the leading and 
subleading terms 
can be written explicitly as 
\begin{align}
\langle \zeta_1 \zeta_2 \rangle &=\frac{ \langle \delta \chi N(\chi)  \rangle^2}{\langle \delta \chi^2\rangle^2}\Sigma (r_{12})\nonumber  \\ &+\frac{\langle \delta \chi^2 (N(\chi) -\bar N) \rangle^2}{2 \langle \delta \chi^2 \rangle^4} \Sigma (r_{12})^2 + ...
\label{subleading}
\end{align}
where $\bar N = \langle N \rangle$ and one can compare the magnitude of the two terms for a given model. 

An alternative approach would 
be to evaluate 
the full expression, Eq.~(\ref{corrFull}) (specialising, for example, to the  two-point function) which always 
remains valid, and compare with the results of the expansion method. 
To do so for a full range of $r_{12}$ would of course negate the advantage of using the 
expansion in the first place, but one could do so for a single representative value of $r_{12}$. 
In the next section when we study simple examples numerically 
we will evaluate the full expression over a range of $r_{12}$, but we note that in more complex cases this 
may not be feasible.

\section{Examples}
\label{examples}

Let us now see our expressions in practice. 
In this paper we restrict ourselves to 
cases in which we already have an 
$N(\vec{\chi})$ function calculated, deferring the Monte Carlo type applications 
discussed in section \ref{monte} to future work.

In addition to a specific $N$ function, for concrete applications, we 
must also specify the statistics of the field fluctuations $\delta \chi^I(\mathbf{x})$.
In order to do so, at this point we specialise to 
uncoupled Gaussian perturbations, with scale invariant power spectrum, such that
\begin{equation}
\label{four}
\Sigma^{IJ} (k) = \delta^{IJ} \frac{P_0}{k^3} \,,
\end{equation}
where $P_0$ is a constant. Moreover, in the examples 
we present we will mainly assume that only the perturbations from 
one field contribute significantly to $\zeta$, and therefore 
we can further specialise to $N$ being a 
function of just a single field.

With our convention for the Fourier Transform
\begin{equation}
\label{fourier}
\delta \chi(\mathbf{x}) = \frac{1}{(2 \pi)^3} \int \di^3 k e^{i \mathbf{k}.\mathbf{x}} \delta \chi_{\mathbf{k}}\,,
\end{equation}
it  follows that
\begin{equation}
\label{corrExplicit1}
\Sigma =\langle \delta \chi^2 (\mathbf{x}) \rangle = \frac{1}{(2 \pi)^3} \int \di^3 k P_\chi (k) = \frac{P_0}{2 \pi^2}  \int^{q_{\rm max}}_{L^{-1}} \frac{\di k}{k} \,,
\end{equation}
where $\sim L^{-1}$  is an
IR and  $q_{\rm max}$  a UV cutoff. In this case, the IR cutoff is just the size of the observable universe, in other words, the scale over which $\bar \chi$ is defined.\\ This gives 
\begin{equation}
\label{seven}
\langle \delta \chi^2(\mathbf{x}) \rangle = \frac{P_0}{2 \pi^2} \ln(q_{\rm max}L)\,,
\end{equation}
for the two-point function of field fluctuations evaluated at the same 
spatial position. Physically, the IR cutoff must be close to the size 
of the observable universe so that the average of $\delta \chi(\mathbf{x})$ 
within the observable universe is zero -- to be consistent with our initial definition 
of $\delta \chi(\mathbf{x}) = \chi(\mathbf{x}) - \bar\chi$.

Next, consider the correlation of the field fluctuations 
at two separated positions. In this case one finds
\begin{eqnarray}
\langle \delta \chi(\mathbf{x}_1) \delta  \chi(\mathbf{x}_2) \rangle
\hspace{-.1cm}&=& \hspace{-.1cm} \frac{P_0}{2 \pi^2} \Bigg(\hspace{-1mm} -\mathcal{C}_i \left(\frac{r_{12}}{L} \right) 
+\mathcal{C}_i(q_{\rm max} r_{12}) \nonumber \\
&&  +\,
\frac{\sin(\frac{r_{12}}{L})}{\frac{r_{12}}{L}}  
 - \frac{\sin(q_{\rm max}r_{12})}{q_{\rm max}r_{12}}\Bigg ) \,,
 \label{field2ptR}
\end{eqnarray}
where $\mathcal{C}_i(x)$ is the cosine integral function
\begin{equation}
\label{ten}
\mathcal{C}_i(x) = - \int^{\infty}_x \frac{\cos(t)}{t} dt\,.
\end{equation}
It is in this cross correlation that the
expansion of section \ref{expansions} was made.
We also define the cross-correlation normalised to the variance as
\begin{equation}
\label{nine}
\frac{\Sigma(r_{12})}{\Sigma}=\xi(r_{12})= \frac{ \langle \delta \chi(\mathbf{x}_1) \delta  \chi(\mathbf{x}_2) \rangle}{\langle \delta \chi^2 (\mathbf{x}) \rangle }\,, \end{equation}
which we require to be small for the expansion of the probability distribution to be valid. 

For a purely scale invariant spectrum and for distances much longer than the 
UV cutoff (i.e., $r_{12}\gg q_{\rm max}^{-1}$), the UV cutoff drops out and we have $\xi(r_{12}) \approx \frac{1}{N_*} \ln \big(\frac{L}{r_{12}}\big)$ \cite{Bethke:2013vca}, where $N_* \approx 60$ is the number of 
e-folds before the end of inflation 
that perturbations corresponding to the largest observable scales left the horizon. For observable scales, therefore, 
$\xi(r_{12}) \approx \frac{1}{60} = 0.017$. This ratio is not sufficiently small that we 
can have complete confidence in the expansion method, especially recalling also the limitations 
mentioned in section \ref{limitations}. We expect, however, that it will likely be sufficiently accurate  
in many cases.

\subsection{Analytic examples}

The next step is to specify the $N(\chi)$ function. To begin with, for simplicity and in order to highlight some 
issues, we follow Ref.~\cite{Suyama:2013dqa} and choose the simple analytic functions studied there.

\subsubsection{Sine function}
First we consider a sine function
\begin{equation}
\label{eighteen}
N(\chi) =B \sin \big ( \frac{\chi}{ \lambda} \big )
\end{equation}
 We compute the two-point function of the curvature perturbation, $\langle \zeta_1 \zeta_2 \rangle$, for this example 
in several ways. 

First, we directly integrate
the fully  non-perturbative expression for $\langle \zeta_1 \zeta_2 \rangle$ which arises from Eq.~(\ref{corrFull}); this 
makes use of the joint probability distribution for $\chi_1$ and $\chi_2$. Because of the simple form 
of the analytical function we have taken for $N$, the resulting integration is easily tractable analytically, 
and we denote the result by
$\langle \zeta_1 \zeta_2 \rangle_{{\rm Full}}$.

In section \ref{expandedmethod}, we presented Eq.~(\ref{dNnew}) 
as the result of our expansion method, and later presented 
a simplified expression for $\tilde N_I$ in Eq.~(\ref{NI}). The second way in which we compute (an approximation to)
$\langle \zeta_1 \zeta_2 \rangle$ is therefore to employ these formulae, leading to
\begin{equation}
\label{ten}
\langle \zeta_1 \zeta_2 \rangle_{{\rm Exp}}= \langle \delta \chi N \rangle^2 \frac{\Sigma(r_{12})}{\langle \delta \chi^2 \rangle^2}\,.
\end{equation}

Taking $\bar \chi = 0$ one finds
\begin{equation}
\label{forty}
\langle \zeta_1 \zeta_2\rangle_{{\rm Full}} = B^2 e^{-\frac{\langle \chi^2 \rangle}{\lambda^2}} \sinh{ \Bigg( \frac{\langle  \chi ^2 \rangle}{\lambda^2} \xi(r_{12})} \Bigg )
\end{equation}
and to leading order
\begin{equation}
\label{fortyone}
\langle \zeta_1 \zeta_2\rangle_{{\rm Exp}} = B^2 e^{-\frac{\langle \chi^2 \rangle}{\lambda^2}}  \frac{\langle  \chi^2 \rangle}{\lambda^2} \xi(r_{12})
\end{equation}
which also follows from expanding Eq.~(\ref{forty}).

This example is useful, because it highlights, as was also noted in Ref.~\cite{Suyama:2013dqa}, the possible 
limitation of our expansion methods discussed in section \ref{limitations}. In this case, for 
$\langle \zeta_1 \zeta_2\rangle_{{\rm Exp}}$ to be a good approximation to $\langle \zeta_1 \zeta_2\rangle_{{\rm Full}}$, 
it is insufficient for only $\xi(r_{12}) \ll 1$. We have to impose a more stringent condition, namely $\xi(r_{12})  \langle \chi^2 \rangle = \Sigma(r_{12}) \ll \lambda^2$. One should note that this is still a significant improvement over the  standard $\delta N$ method of making a Taylor expansion of the $N$ function reviewed in section \ref{regular}. $\lambda$ is a measure of the width 
of a feature in the $N$ function, and the requirement for standard $\delta N$ to work is that $\Sigma \ll \lambda^2$, 
while for our expansion method only that  $\Sigma(r_{12}) \ll \lambda^2$ is required, which as we have seen is two orders of 
magnitude less 
stringent.

\subsubsection{Gaussian function}
 
For our second analytic example, we consider the $N(\chi)$ function to be an un-normalised Gaussian
\begin{equation}
\label{eighteen}
N(\chi) = A \frac{e^{\frac{-(\chi -m_1)^2}{2 \sigma_1^2} }}{ \sqrt{2 \pi} \sigma_1} \,
\end{equation}
where $A$, $m_1$ and $\sigma_1$ are constants defining the amplitude, position of the peak and width of the function. 
In Ref.~\cite{Suyama:2013dqa} the authors used a sum of normal distributions with different amplitudes and widths
to represent the spiky $N(\chi)$ function that arises in massless preheating \cite{Chambers:2007se,Chambers:2008gu,Bond:2009xx}.

Without loss of generality we can take $\bar \chi =0$. Here 
we denote the variance of the probability distribution of the field perturbations, $\Sigma$, using 
$\Sigma = \sigma^2$, and doing so
we find\begin{equation}
\label{nineteen1}
\begin{split}
\langle \zeta_1 \zeta_2\rangle_{{\rm Full}} \,= \,& A^2\frac{e^{-\frac{m_1^2}{\sigma^2+\sigma_1^2+\Sigma(r_{12})}} }{2 \pi \sqrt{(\sigma^2 + \sigma_1^2)^2 - \Sigma(r_{12})^2}}  \\&-  A^2\frac {e^{-\frac{m_1^2}{\sigma^2 + \sigma_1^2}}} {2 \pi (\sigma^2 + \sigma_1^2) },
\end{split}
\end{equation}
 and to leading and subleading order, from Eq.~(\ref{subleading}), we have
\begin{equation}
\begin{split}
\label{nineteen}
\langle \zeta_1 \zeta_2\rangle_{{\rm Exp}} = & A^2 \frac{e^{-\frac{m_1^2}{\sigma^2 + \sigma_1^2}}m_1^2 \Sigma(r_{12})}{2 \pi (\sigma^2 + \sigma_1^2)^3} \\+ & A^2\frac{e^{-\frac{m_1^2}{\sigma^2 + \sigma_1^2}}(-m_1^2 + \sigma^2  + \sigma_1^2)^2 (\Sigma(r_{12}))^2 }{4 \pi (\sigma^2 + \sigma_1^2)^5 }\,,
\end{split}
\end{equation}
which also follows from expanding Eq.~(\ref{nineteen1}).

The ratio of the subleading term to the leading term is 
\be
{\rm ratio} =  \frac{\Sigma (r_{12}) (-m_1^2  + \sigma^2 + \sigma_1^2)^2    } {2 m_1^2  (\sigma^2 + \sigma_1^2)^2 } \,\nonumber.
 \ee
 We wish to understand when this is small, and hence when our expansion method 
 can be trusted. 
 Assuming $\sigma_1 \leq \sigma$ (the $N$ function is of a similar width or narrower  than the 
 distribution  of field perturbations), 
 the condition required for the ratio to be small becomes $ \Sigma (r_{12}) \ll  m_1^2 \sigma^4  /(-m_1^2  +  \sigma^2 )^2   $. For fixed $\sigma$, there 
 is then both a lower and an upper limit on $m_1$ in order for this condition to be satisfied.
 This makes sense since if $m_1$ is too small, which in this case means $m_1 \ll \sigma$ the $N$ function becomes  
 close to even. While if  $m_1 \gg \sigma$ 
 the $N$ function is sampled only by the tail of the probability 
 distribution, and one would not expect the expansion to be be accurate. A representative case is $m_1\sim {\cal O}( \sigma)$, 
leading to $ \Sigma (r_{12}) \ll \sigma^2$, which is the condition we assumed to make our original expansion.
 
The other case is where  
$\sigma_1 \geq \sigma$. In this case the distribution is now narrower than the $N$ function, and 
the ratio implies we must have $ \Sigma (r_{12}) \ll  m_1^2 \sigma_1^4 /(-m_1^2  +\sigma_1^2)^2$. In this 
case the ratio can also be satisfied as long as $m_1$ is not too small or too large, which in this case means neither 
$m_1 \ll \sigma$ nor $m_1 \gg \sigma$.
In the representative case of $m_1\sim {\cal O}(\sigma_1)$, the condition reduces to  
$\Sigma (r_{12}) \ll  \sigma_1^2$, which is weak given that  $\sigma_1 > \sigma$. 
We would expect standard $\delta N$ to work in the case
($\sigma \ll \sigma_1$), but here, as for the sinusoidal case, we have relaxed that criteria.

\subsubsection{Lessons}
It is also important to note that in all the cases above, the expansion fails because 
the leading contribution to the two-point function of $\zeta$ itself becomes very small. 
In the second example, if the $N$ function was made up of a series of spikes (as is the case where the result of massless pre-heating is parametrised), even if the expansion failed for some members of the series, 
the overall value for the leading term would be dominated by members of the series for which $m_1$ does not fall outside 
the allowed range, leading to an accurate overall result. This also gives us hope that for a realistic $N$ function, calculated, for example, from lattice simulations the expansion method we advocate will be accurate.

It seems therefore that there are two regimes in which the method has a good chance of working.  One either requires that 
 $\Sigma(r_{12})^{1/2}$ is smaller than the scale on which the $N$ function is structured, or that $\Sigma^{1/2}$, is much larger 
 than the scale on which the $N$ function is structured (and so the structure is averaged over, assuming the average is not close to zero). In intermediate cases the method seems to fail.  Overall, however, the message of these 
two analytic examples is that it is crucial to check for the validity of the approximation on a case by case basis.

\subsection{A Non-analytic example}
\begin{figure}
\centerline{\includegraphics[angle=0,width=95mm]{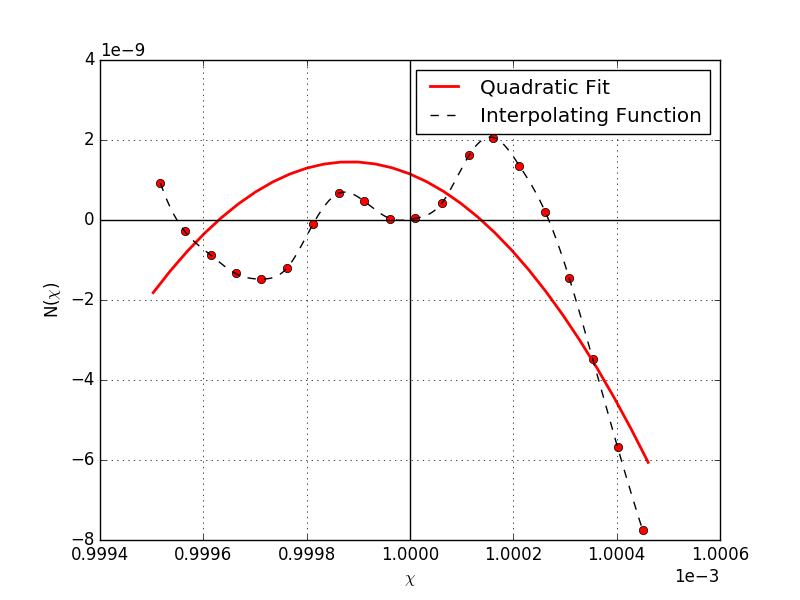}}  
\caption{An example of a realistic mapping obtained from lattice field theory simulations and centered around $\bar \chi = 0.001$ \cite{Chambers:2009ki}. Red dots are the data points, the black dashed line shows the interpolating function and the solid red line represents a quadratic fit to the data points. We will use the interpolating function for our regular $\delta$N analysis.} 
\label{Datapoints}
\vspace*{0pt}
\end{figure}

Next we turn to a more 
realistic example. Although almost all the analysis of the curvaton scenario is based on the assumption of a perturbative curvaton decay, it is possible for the curvaton to decay through a non-perturbative process analogous to inflationary preheating \cite{Enqvist:2008be,BasteroGil:2003tj}. For our example, we consider the $N(\chi)$ function presented in Fig.~3 of Ref.~\cite{Chambers:2009ki}, which 
was generated from a resonant curvaton decay scenario using classical lattice field theory simulations \cite{Khlebnikov:1996mc,Prokopec:1996rr}.
The system consists of three fields: an inflaton, curvaton and a third light 
 field, $\chi$.  The curvaton field decays into particles of $\chi$ via parametric resonance \cite{Traschen:1990sw,Kofman:1994rk,Kofman:1997yn}.  
 The authors considered only the contribution of perturbations from the $\chi$ field to 
  $\zeta$, and so $N$ is a function only of this field. 
 In order 
 to perform the integrations necessary to study this model, we 
 construct an interpolating function to approximate $N(\chi)$ given the data points 
 presented in Ref.~\cite{Chambers:2009ki}. We present the data points and the interpolating function 
 in Fig.~\ref{Datapoints}.

In this section we will again 
compute the two-point function of the curvature perturbation in real space,  $\langle \zeta_1 \zeta_2 \rangle_{\rm Full}$, 
from 
Eq.~(\ref{corrFull}) as described above, 
and then using our expansion method (retaining only the leading term) we will calculate $\langle \zeta_1 \zeta_2 \rangle_{\rm Exp}$. 
This time both must be computed numerically, and 
this means we have to fix the various parameters which enter the expression presented at the start of section \ref{examples}, in 
particular,  the IR cutoff $L^{-1}$ and the UV cutoff $q_{\rm max}$. We do so by assuming that perturbations 
which exited the horizon $60$ e-folds before the end of inflation correspond to the largest observable scales today.
We associate the largest observable scale today with $L$, and include in the calculation all shorter modes which exit the horizon 
until the end of inflation. Taking the scale of the shortest modes to be $r_{\rm min}$, it then follows that $L = e^{60} \times r_{\rm min} \approx 10^{26} r_{\rm min}$. The UV cutoff, defined as $q_{\rm max}=\frac{2 \pi}{r_{\rm min}} $.

We will also compute the power spectrum in Fourier space, and the methods we use for this are discussed in the 
next subsection. Since the scales constrained by CMB anisotropy data correspond to 
the modes which exited during roughly 4 e-folds of inflation, when presenting our results the range of $k$ values we will interested in range from
$\frac{2 \pi}{L}$ to $e^4 \times \frac{2 \pi}{ L}$, i.e., from the horizon size today 
down to about $e^4$ times smaller than the horizon size. 

In addition to the full and expanded expressions, we will also plot the results for the power spectrum that one attains from the 
regular $\delta N$ method, calculating the derivatives of $N$ locally at our choice of the value of $\bar \chi$. 
Finally using our expansion method, we will also calculate the reduced bispectrum $\fnl$ for this model, comparing with the results 
which would be obtained from regular $\delta N$. 

\begin{figure}[]
\centerline{\includegraphics[angle=0,width=90mm]{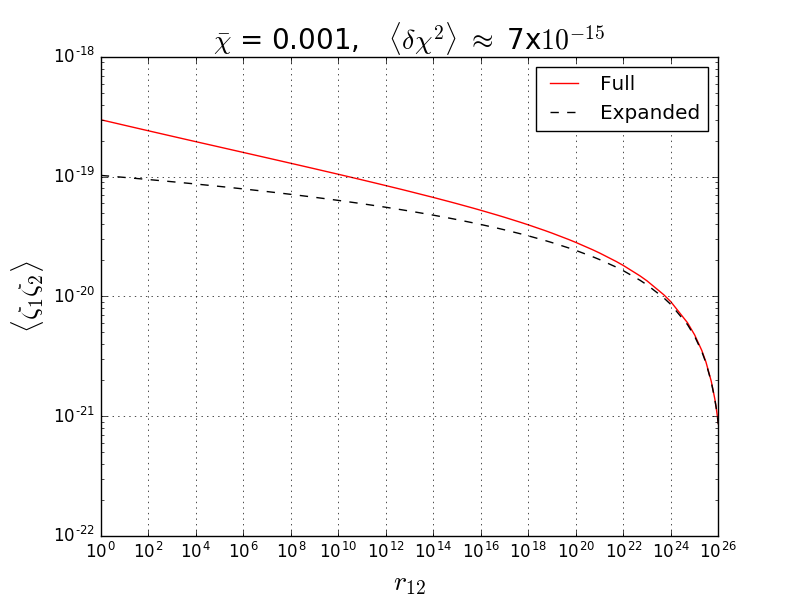}}  
\caption{Case 1: Correlation function of $\zeta(\chi)$ for one realization with $\bar \chi=0.001$ and  $\langle \delta \chi^2 \rangle \approx 7 \times$$10^{-15}$ on a Log-Log plot. The exact correlation function (`Full') is calculated from Eq. (11). The approximated correlation function (`Expanded') is  given by Eq. (16). As expected, the approximated correlation function becomes progressively worse on shorter scales.} 
\label{realSpaceNum1}
\centerline{\includegraphics[angle=0,width=90mm]{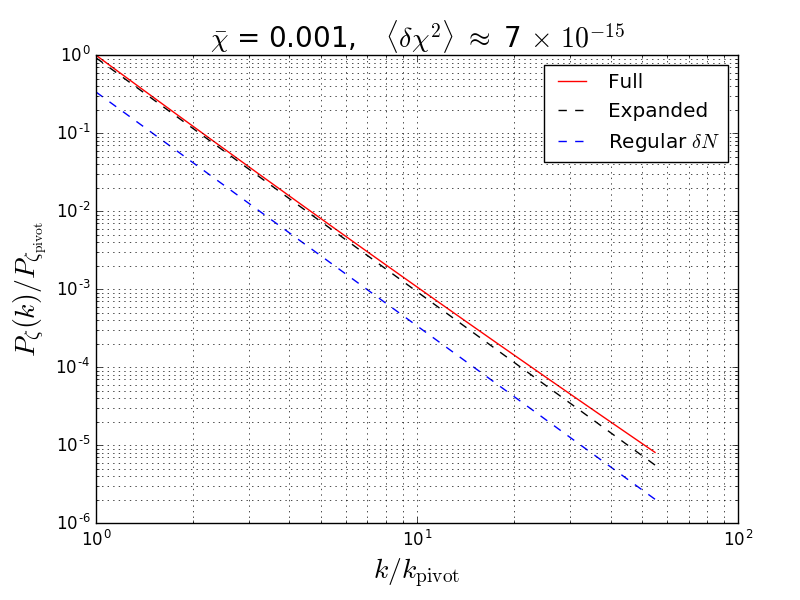}}  
\caption{Case 1: Log-Log plots of the power spectrum of $\zeta$, calculated using the full, expansion and regular $\delta$N methods respectively.} 
\label{fSpaceNum1}
\end{figure}
 \begin{figure}[]
\centerline{\includegraphics[angle=0,width=90mm]{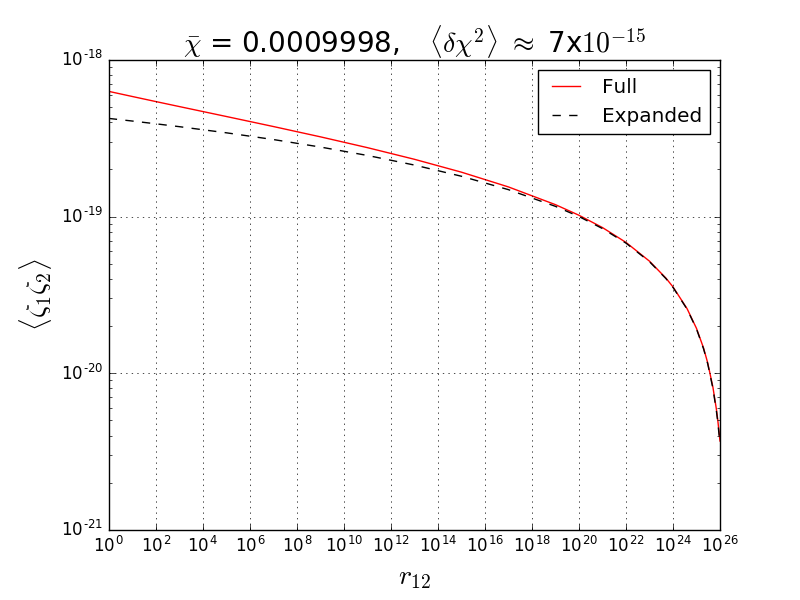}}  
\caption{Case 2: Here, we plot the correlation function of $\zeta(\chi)$ for one realization with $\bar \chi=0.0009998$ and  $\langle \delta \chi^2 \rangle \approx 7 \times$$10^{-15}$ on a Log-Log plot.} 
\label{realSpaceNum2}
\centerline{\includegraphics[angle=0,width=90mm]{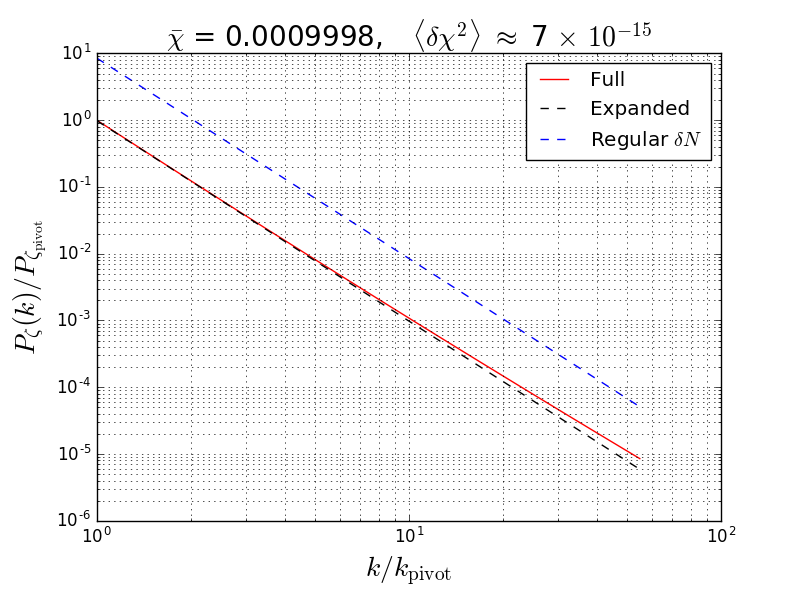}}  
\caption{Case 2: Log-Log plots of the power spectrum of $\zeta$, calculated using the full, expansion and regular $\delta$N methods respectively.} 
\label{fSpaceNum2}
\end{figure}
\subsubsection{The Power Spectrum}
Our expansion method, Eq.~(\ref{dNnew}) allows us to pass directly to Fourier space and to write 
the power spectrum as $P_\zeta(k)_{\rm Exp} \approx \tilde N_I \tilde N_J \Sigma^{IJ}(k)$ 
where $\Sigma^{IJ} (k) = \delta^{IJ} \frac{P_0}{k^3}$. 
However, if one wishes to work with the fully non-perturbative $\langle \zeta_1 \zeta_2 \rangle_{{\rm Full}}$, one 
needs to Fourier transform the real space two-point function of $\zeta$. 
The route we take to achieving this is as follows. First we define
\begin{equation}
\label{twentythree}
\begin{split}
{\mathscr{F}}\left [\, \langle\zeta_1 \zeta_2\rangle_{{\rm Full}}\,\right]&=\langle\zeta_\mathbf{{k_1}} \zeta_\mathbf{{k_2}} \rangle_{{\rm Full}} \\&= (2 \pi)^3 \delta^3 (\mathbf{k_1} + \mathbf{k_2}) P_{\zeta} (k_1)_{{\rm Full}}\,.
\end{split}
\end{equation}
Then given that the 
 two-point function is always some function of $r_{12}=\lvert \mathbf{r_1}-\mathbf{r_2} \rvert$, we 
 define $\langle\zeta_1  \zeta_2 \rangle_{{\rm Full}}= A(r_{12})$ and note
\begin{equation}
\label{twentyfour}
{\mathscr{F}}\left [  A(r_{12}) \right ] \hspace{-.05cm }= \hspace{-.05cm}\int_{-\infty}^{\infty} \hspace{-.05cm}  \mathrm{d}^3\mathbf{r_1}  \int_{-\infty}^{\infty}  \hspace{-.05cm}\mathrm{d}^3 \mathbf{r_2}  A(r_{12}) e^{-i\mathbf{k_1}.\mathbf{r_1}} e^{-i\mathbf{k_2}.\mathbf{r_2}}\,.
\end{equation}
By making a change of variables from ${\bf r}_1$ to ${\bf r}_{12}$ we can pull out an delta function to write
\begin{equation}
{\mathscr{F}} \left[  A(r_{12}) \right ]\hspace{-.05cm}= \hspace{-.05cm}(2 \pi)^3  \delta^3 (\mathbf{k_1}+\mathbf{k_2})  \int_{-\infty}^{\infty} \mathrm{d}^3 \hspace{-.05cm}\mathbf{r_{12}}   A(r_{12}) e^{-i\mathbf{k_1}.\mathbf{r_{12}}}\,.
\label{intermediateFT}
\end{equation}
Therefore we arrive at the expression
\begin{equation}
\label{twentysix}
P_{\zeta}(k_1)_{{\rm Full}}=  \int_{-\infty}^{\infty} \mathrm{d}^3\mathbf{r_{12}}  \; A(r_{12}) \;e^{-i\mathbf{k_1}.\mathbf{r_{12}}}\,,
\end{equation}
which on moving to spherical polar coordinates leads to the one dimensional integral
\begin{equation}
\label{twentyseven}
P_{\zeta}(k_1)_{{\rm Full}}=  4 \pi \int_{0}^{\infty} \mathrm{d} r_{12} \;  r_{12}^2 A(r_{12}) \; \frac{\sin(k_1 r_{12})}{k_1 r_{12}} \,.
\end{equation}

To evaluate the power spectrum, therefore, one possibility 
is to first use the $N(\chi)$ function to calculate $A(r_{12})$ for a range of values of $r_{12}$, and then 
to perform this one dimensional integration. Rather than sampling $A(r_{12})$ at all positions needed by an integration algorithm,
one could fit $A(r_{12})$ with an interpolating function. A problem that arises, however, is that the integral is sensitive to the value of the integrand even for $r_{12} \gg L$. A second issue is that the integrand is highly oscillatory.  
These issues meant we couldn't get accurate results using this strategy. 
An alternative is to evaluate instead Eq.~(\ref{twentysix}), using a fast (discrete) Fourier transform. Although this is effectively a three dimensional 
integral, the speed of the algorithm involved means it is more tractable than integrating Eq.~(\ref{twentyseven}). To avoid aliasing, 
we must sample $A(r_{12})$ with a small enough uniform intervals such that the sampling frequency is at least twice the highest frequency contained in the signal. In this case, the highest frequency that we're interested in is $e^4 \times \frac{2 \pi}{ L}$ and we always ensure this 
criteria is easily met. We must also ensure that the lowest frequency sampled is at least an order of magnitude smaller than $\frac{2 \pi}{ L}$. Even when these constraints are met, the results of the Fourier transform will have a number of spurious points. In order to present a clean plot, therefore, we fit the data in log 
space to a polynomial. Finally we plot this fitted function. As a test that we are sampling the correct range and the method 
is working, we first applied it to a sampled version of Eq.~(\ref{field2ptR}), 
to ensure we recovered Eq.~(\ref{four}) with precision.

\begin{figure}[]
\centerline{\includegraphics[angle=0,width=90mm]{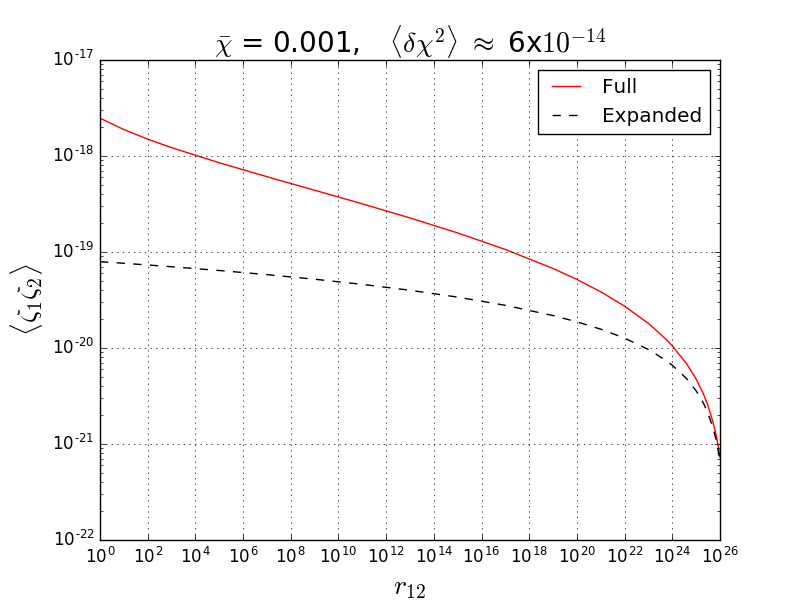}}  
\caption{Case 3: Here, we plot the correlation function of $\zeta(\chi)$ for one realization with $\bar \chi=0.001$ and  $\langle  \delta \chi^2 \rangle \approx 6 \times$$10^{-14}$. The approximated correlation function is  worse in this case because the shorter tail distribution `sees' less of the mapping.} 
\label{realSpaceNum3}
\centerline{\includegraphics[angle=0,width=90mm]{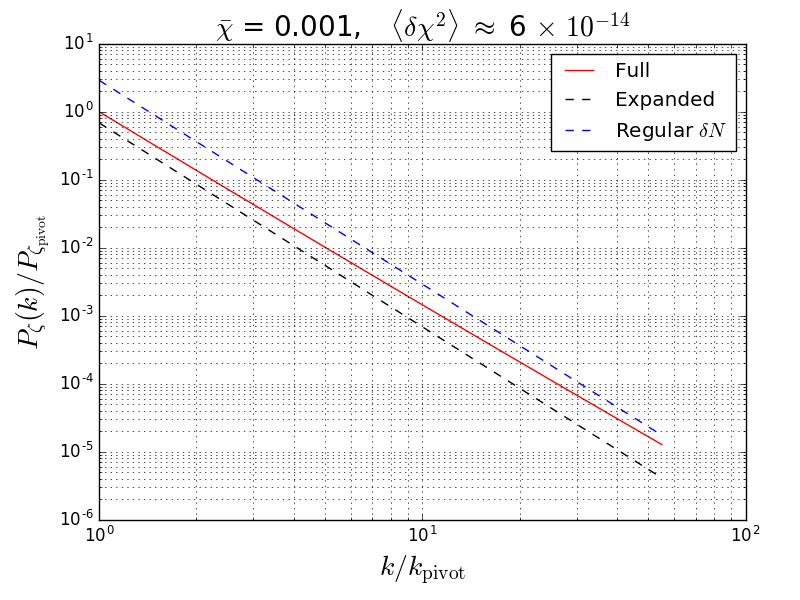}}  
\caption{Case 3: Log-Log plots of the power spectrum of $\zeta$, calculated using the full, expansion and regular $\delta$N methods respectively. The effect of the shorter tail distribution is also reflected in the difference between the `Full' and `Expanded' power spectra. } 
\label{fSpaceNum3}
\end{figure}

\subsubsection{Three cases}

We perform our analysis for three cases and present our analysis of the 
two-point function and the power spectrum in Figs.~\ref{realSpaceNum1}-\ref{fSpaceNum3}. The cases we 
consider are
\begin{enumerate}
 \item $\bar \chi=0.001$ and  $\langle \delta \chi^2 \rangle \approx 7 \times $$10^{-15}$
 \item $\bar \chi=0.0009998$ and  $\langle \delta \chi^2 \rangle \approx 7 \times$$10^{-15}$
 \item $\bar \chi=0.001$ and  $\langle \delta \chi^2 \rangle \approx 6 \times$$10^{-14}$
\end{enumerate}

For the power spectrum, we plot $P_{\zeta}(k)/P_{\zeta_{\rm pivot}}$ against $k/k_{\rm pivot}$.
 We arbitrarily choose $k_{\rm pivot} = \frac{2 \pi}{L}$. We also fix $P_{\zeta_{\rm pivot}}$ to be $P_{\zeta_{\rm Full}}\rvert_{k=k_{\rm pivot}}$ for all three (`Full', `Expanded', `Regular $\delta$N')
methods for easy comparison; otherwise all three lines will lie on top of each other initially 
as dividing by each of their corresponding pivot value of $P_{\zeta}$ will force them to start at the same point. 

In all cases we see that the expansion method is a much better approximation to the fully non-perturbative method than 
regular $\delta N$, and in two of the cases does a good job at recovering the amplitude and initial scale dependence of the power spectrum. In the third case however, we can see the method is breaking down even for the largest scales. 

In all three cases, we see that the `Expanded' power spectrum either matches or is smaller that the `Full' power spectrum on all scales while the `Regular' power spectrum can be smaller or larger than the `Full' answer, depending on the value of $\bar \chi$ and $\langle  \delta \chi^2 \rangle$.

\subsubsection{The Reduced Bispectrum}

First, we calculate the reduced bispectrum using regular $\delta N$. For case 1, $f_{\rm NL}$ is $\mathcal{O}(10^{10})$, $f_{\rm NL}$ is negative and $\mathcal{O}(10^{7})$ for case 2 and finally, $f_{\rm NL}$ is $\mathcal{O}(10^{10})$ for case 3. 
Using Eq. (20), i.e., using the expansion method we can  also calculate the reduced bispectrum in each case. We find that $f_{\rm NL}$ is enormous for all three cases:
 $f_{\rm NL}$ is $\mathcal{O}(10^{9})$, $\mathcal{O}(10^{8})$ and $\mathcal{O}(10^{10})$ for case 1, 2 and 3 respectively. 
 This is to be expected since in all cases the higher order terms in the non-perturbative $\delta N$ expansion 
 are relatively large (since by eye one can see the full line deviate from the expanded line plotted using only the leading term). 
 
 However, we also find that the amplitude of the curvature perturbation  for these specific examples 
is too small to explain the observed amplitude: $\mathcal{O}(10^{-20})$ , $\mathcal{O}(10^{-19})$ and $\mathcal{O}(10^{-20})$
for case 1, 2 and 3 respectively. It is likely this can be altered by changing $\langle \delta \chi \delta \chi \rangle$. But given the $N(\chi)$ function 
we began with, we are limited to assuming $\langle \delta \chi \delta \chi \rangle^{1/2}$ is much smaller than the range of $\chi$ over which the 
$N$ function has been calculated. Ultimately $\langle \delta \chi \delta \chi \rangle^{1/2}$ is fixed by the energy scale of inflation, but 
unlike in the usual approach we can't account for the effect of changing this energy scale after calculating the derivatives 
of $N$, because the non-perturbative nature of the calculation means the non-perturbation $\delta N$ coefficients are 
affected by $\langle \delta \chi \delta \chi \rangle^{1/2}$.

In terms of the parameters we are working with, therefore, in order to agree with observation we would require that the total curvature perturbation is a mixture of the subdominant component that we have and another dominant component. Taking the observed amplitude to be $10^{-9}$ \cite{Ade:2015lrj} and taking the dominant component to be the standard adiabatic Gaussian perturbation from the inflaton, this mixture dilutes the non-Gaussianity of the total curvature perturbation and as a result,
 $f_{\rm NL}$ becomes $\mathcal{O}(10^{-13})$, $\mathcal{O}(10^{-12})$
and $\mathcal{O}(10^{-12})$ for case 1, 2 and 3 respectively (assuming the inflaton contribution is completely Gaussian) which is far below the observational sensitivity \cite{Ade:2015ava}.

\section{Conclusion}
\label{conclusion}

In the regular $\delta$N formalism, the mapping between the curvature perturbation $\zeta$ and the scalar field(s) 
fluctuations is approximated by a Taylor expansion in the fields. This standard technique fails in some cases. 
Examples include the massless preheating model and the non-perturbative curvaton decay model we revisited in 
the examples section of this work. In this work, we discuss how  
to calculate correlation functions of $\zeta$ when the mapping is an arbitrary function of the scalar field(s) 
without making a Taylor expansion. This entails integrating the full probability distribution of the field 
fluctuations against copies of the $N$ function relating e-folds to initial field values (`Non-perturbative $\delta$N formalism').
We discuss 
how to calculate results using a `Full' (not approximated) implementation of this formalism, but show 
that this can be convoluted in practice. For observationally 
relevant scales the task can be made simpler 
using an expansion method. This leads to a set of expressions for 
observable quantities in terms of non-perturbative $\delta N$ coefficients analogous to the usual $\delta N$ coefficients (`Expanded').
We argue that the validity of the expansion method must be tested on a 
case by case basis and suggest ways to do this, but show that at least 
in the realistic example we consider it leads to a marked improvement over regular $\delta N$, and 
can approximate well the full result.

Our results are closely related to the work of 
Suyama and Yokoyama \cite{Suyama:2013dqa} and Bethke {\it et al.} (\cite{Bethke:2013aba}, \cite{Bethke:2013vca}), but we diverge from their work in a number of ways. 
First we show how to incorporate the perturbations from $n$ fields whose initial probability 
distribution need not be precisely Gaussian, and we present our expressions in an alternative way to those authors, which is 
more suitable for numerical analysis. The expressions are, as we discuss in section \ref{simple}, particularly well suited 
to settings in which a Monte Carlo approach can be advantageous. We intend to employ our results in this 
setting in forthcoming work, directly utilising lattice simulations.

It might seem odd at first that we can use the separate universe approach and 
information from lattice simulations, which simulate only very short scales, 
to infer information about perturbations on observable scales. This works, however, because the  non-pertubative method 
works in real space initially, and at first calculates quantities such as $\langle \zeta(\mathbf{x}) \zeta(\mathbf{y})\rangle$ 
without coarse-graining. As long as the simulations are of regions larger than the horizon during reheating, therefore,
there is then no barrier to using this method together with $\delta N$ to calculate  $\langle \zeta(\mathbf{x}) \zeta(\mathbf{y})\rangle$.
This is not directly observable, since it includes information about all scales which aren't observable. After calculating 
it, however, we can take its Fourier transform and consider the Fourier modes over the range of observable scales (or equivalently coarse-grain the real space result on these scales) to  compare with observations. The method we present, therefore, represents 
a unique opportunity to extract for the first time 
observable predictions for the curvature perturbation directly from lattice simulations.

\section*{Acknowledgements} 
DJM is supported by a Royal Society University Research Fellowship and SVI acknowledges the support of the STFC grant ST/M503733/1. AR is supported by the STFC grant ST/P000762/1.

\bibliography{mybib}

\end{document}